\documentclass[journal=jacsat,manuscript=article]{achemso}

\usepackage[version=3]{mhchem} 

\author{Rodrigo Becerra Silva}
\author{Jay Huang}
\author{Bob Minyu Wang}
\author{Ziyi Song}
\author{Henry Clark Travaglini}
\author{Dong Yu}
\email{yu@physics.ucdavis.edu}
\affiliation[University of California, Davis]
{Department of Physics and Astronomy, University of California, Davis}

\title[An \textsf{achemso} demo]
  {Super-diffusion of Photoexcited Carriers in Topological Insulator Nanoribbons}


\begin{document}

\begin{tocentry}
\includegraphics[scale=1]{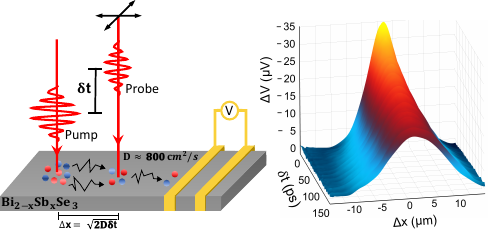}

\end{tocentry}


\begin{abstract}
Understanding the ultrafast dynamics and transport of photoexcited carriers in topological insulators is crucial for the optical manipulation of spins and may shed light on the nature of topological excitons. Here we investigate bulk-insulating Sb-doped $\mathrm{Bi_2Se_3}$ nanoribbons via ultrafast transient photovoltage microscopy. The probe-pulse-induced photovoltage is substantially suppressed by a pump pulse. Recovery time increases from 50 to 1600 picoseconds as the pump fluence increases. We found that the diffusivity of photoexcited carriers increases significantly at lower carrier concentrations, up to 800 cm$^2$/s at 21 K, two to three orders of magnitude higher than that of band-edge carriers. Remarkably, the photoexcited carriers travel up to 10 $\mu$m for hundreds of picoseconds at this high diffusivity. The diffusivity peaks in intrinsic devices and is reduced at high temperatures. We discuss the possible mechanisms of long-ranged super-diffusion in the frames of hot carriers and exciton condensation.
\end{abstract}


\newpage

The surface states of topological insulators (TIs) exhibit unusual Dirac dispersion relation, symmetry-protected transport against external perturbation, and helical spin texture. These features open a new venue for optical control of spin polarization  \cite{mciver2012control,braun2016ultrafast, kastl2015ultrafast} and Floquet quantum states \cite{wang2013observation}. Furthermore, non-equilibrium excitons have been experimentally demonstrated in both graphene \cite{liu2017quantum,li2017excitonic,ju2017tunable} and TIs \cite{kung2019observation, mori2023spin}. Excitons are spin integer particles and obey Bose-Einstein statistics. The realization of Bose-Einstein condensation requires that the wavefunctions of particles overlap. Dirac materials with vanishing effective mass may host excitons of long de Broglie wavelengths and high critical temperature ($T_c$). TIs have been theoretically predicted as a promising platform for achieving high-$T_c$ exciton condensation \cite{triola2017excitonic, pertsova2018excitonic, wang2019prediction}. We have recently observed highly dissipationless transport of photogenerated carriers in TIs, indicating the formation of exciton condensates \cite{hou2019millimetre}. The coherent quantum state of exciton condensates can provide an alternative approach to superconductors for quantum computing that can operate at higher temperatures. Although free fermions have been extensively studied in TIs, much less work is carried out to understand topological excitons. The transport of photoexcited carriers in TIs at ultrafast timescales is not yet understood. Prior experimental efforts in studying photoexcitation in TIs include time-resolved angle-resolved photoemission spectroscopy (tr-ARPES) \cite{sobota2012ultrafast,mori2023spin}, transient reflectance (TR) \cite{kumar2011spatially,gross2021nanosecond}, transient absorption (TA) \cite{glinka2021clarifying,zhou2022transient}, and THz emission methods \cite{braun2016ultrafast,lee2022ultrafast}. Most of these works used photoexcitation over a large area, preventing the extraction of spatially resolved information. 

In this letter, we employ ultrafast transient photovoltage (TPV) using a focused laser with a 1 $\mu$m radius, crucial to studying the transport of photoexcited carriers. TPV is complementary to all-optical pump-probe techniques, where the probe-induced photovoltage is modified by the pump-generated carriers. The characteristic delay time between the pump and probe can be used to study the photoexcitation relaxation dynamics. Recently, TPV and transient photocurrent (TPC) have gained popularity in investigating nanoscale and low dimensional materials~\cite{zeng2023ultrafast}, including determining photoresponse time in graphene and transition metal dichalcogenide (TMD) devices~\cite{sun2012ultrafast, massicotte2016picosecond}, imaging ultrafast carrier transport~\cite{son2014imaging}, and detecting dark excitons~\cite{yagodkin2023probing}. TPV is particularly suitable for studying TIs, as most TIs have narrow bulk bandgaps that make it difficult to access surface states using pure optical methods. We focus on $\mathrm{Bi_2Se_3}$, a prototypical three-dimensional (3D) TI with a relatively large bulk bandgap (approximately 0.3 eV) \cite{Xia2009bandgap}. The addition of Sb in $\mathrm{Bi_2Se_3}$ shifts the Fermi level within the bulk bandgap and slows down the carrier recombination lifetime to nanoseconds as demonstrated by TR measurements \cite{gross2021nanosecond}. $\mathrm{Bi_{2-x}Sb_xSe_3}$ nanoribbons exhibit highly non-local photocurrent generation, suggesting the formation of superfluid-like exciton condensates \cite{hou2019millimetre}. Here we aim to reveal the mechanism of non-local photocurrent generation by using spatiotemporally resolved photovoltage measurements.

\begin{figure}
    \centering
    \includegraphics[width=1\linewidth]{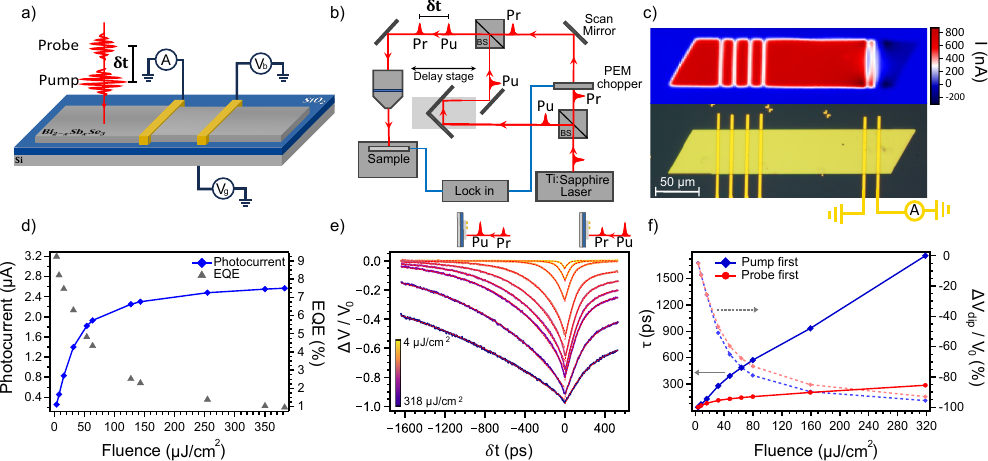}
    \caption{TPV measurements of a $\mathrm{Bi_{2-x}Sb_xSe_3}$ nanoribbon FET at 13 K. (a) Cartoon of the TPV measurement where consecutive laser pulses with delay time $\delta t$ are focused out of the device channel. (b) Schematic of the experimental setup. (c) Top: Photocurrent map obtained at 12 K by scanning the laser with 690 nm wavelength and 6.1 $\mu$J cm$^{-2}$ fluence. Bottom: Device's optical image. (d) Fluence-dependent photocurrent and EQE with the laser focused outside the device channel.  (e) TPV as a function of delay time at various pump fluences. $\delta t < 0$ means the pump pulse arrives earlier than the probe. The probe fluence is 4 $\mu$J cm$^{-2}$. The dashed lines are single exponential fittings. (f) Pump fluence-dependent recovery time $\tau$ extracted from (e) and photovoltage percentage change at zero delay. }
    \label{fig:intro}
\end{figure}

Single crystalline $\mathrm{Bi_{2-x}Sb_xSe_3}$ nanoribbons were grown by chemical vapor deposition (CVD) \cite{kong2010topological,hou2019millimetre}. The atomic percentage of Sb was 6 - 8\% as measured by energy dispersive X-ray spectroscopy (EDS) (Figure S1 in Supporting Information). The nanoribbons were typically 100-200 nm thick, confirmed by atomic force microscopy (AFM) (Figure S2). Field effect transistors (FETs) incorporating single nanoribbons exhibited linear current-voltage characteristics with negligible contact resistance. The conductance decreased at lower temperatures, indicating a semiconducting behavior (Figure S2). The field effect characteristics vary from device to device. Among the 8 devices measured, 3 exhibited ambipolar conduction, while the rest showed $n$-type behavior. The electron mobility ranges from 2.6 to 330 cm$^2$/Vs and density from $2.5\times10^{18}$ to $2.6\times10^{19}$ cm$^{-3}$ (Table S1 in SI). We note that the field effect measurement underestimates these electron mobility values because the calculation assumes that the gate-induced carriers are uniformly distributed across the nanoribbons, which is unlikely due to their relatively large thicknesses. 

We present the photocurrent mapping results for a typical device (D1) below. Above 200 K, the photocurrent was only large when the photoexcitation was close to the contact. However, at lower temperatures, the photocurrent decay length increased sharply (Figure S3). The photocurrent was essentially independent of the laser position at 12 K [Figure \ref{fig:intro}(c)], consistent with our previous report \cite{hou2019millimetre}. The external quantum efficiency (EQE), defined by the ratio of collected electrons to incident photons, reached about 10\% at low laser fluence [Figure \ref{fig:intro}(d)]. Such efficient non-local photocurrent generation suggests that photoexcited carriers can travel across hundreds of $\mu$m before recombination. The photocurrent showed a sublinear dependence on fluence [Figure \ref{fig:intro}(d)], indicating reduced EQE at higher fluence.

The open-circuit photovoltage was then measured as a function of the delay time between the pump and probe pulses. The photovoltage is reduced at zero delay by up to 100\% at high fluence [Figure \ref{fig:intro}(e)]. Such a large change cannot be attributed to the sample reflectance change, which is less than 1\% under a similar pump fluence \cite{gross2021nanosecond}. Instead, we attribute the photovoltage reduction to faster carrier recombination at higher photoexcited carrier density. Previous work \cite{gross2021nanosecond} shows that the bimolecular term dominates the recombination process in our nanoribbons under the experimental conditions applied. Consequently, the higher carrier density produced when the two pulses excite simultaneously leads to more recombination loss over the laser spot, which may substantially lower the carrier collection efficiency. 

The photovoltage suppression is recovered after a characteristic delay time [Figure \ref{fig:intro}(e)]. The TPV curves fit well to an exponential function [$V = V_0 + \Delta V_{dip} \exp(-|\delta t|/\tau)$] for both positive and negative delay times, from which a recovery time $\tau$ can be extracted. On the pump first side ($\delta t<0$), $\tau$ increases from 50 to 1600 ps as the pump fluence increases [Figure \ref{fig:intro}(f)]. On the probe first side ($\delta t>0$), $\tau$ is shorter but also increases with higher fluence. $\tau$ values and the trend of fluence dependence are insensitive to the laser position, even when the laser is focused close to the contacts (Figure S4). TPV is also insensitive to the excitation energy in the tested range of 1.2-1.8 eV (Figure S5). 

At first glance, the slower recovery at higher fluence appears counterintuitive, since the photoexcited carriers recombine faster at higher fluence as observed in TR measurements \cite{gross2021nanosecond}. It should be noted that the laser spot diameter used in the TR work was 40 $\mu$m, while in this work the diameter was about 2 $\mu$m. Hence, under our experimental conditions, photovoltage recovery occurs when the carriers, photoexcited by the first pulse, have time to escape the laser spot before the second pulse arrives. The recombination time has been previously determined to be up to a few ns \cite{gross2021nanosecond}, much longer than the observed photovoltage recovery time at low fluence, indicating the carrier escape likely dominates the recovery process. 
\begin{figure}
    \centering
    \includegraphics[width=0.65\linewidth]{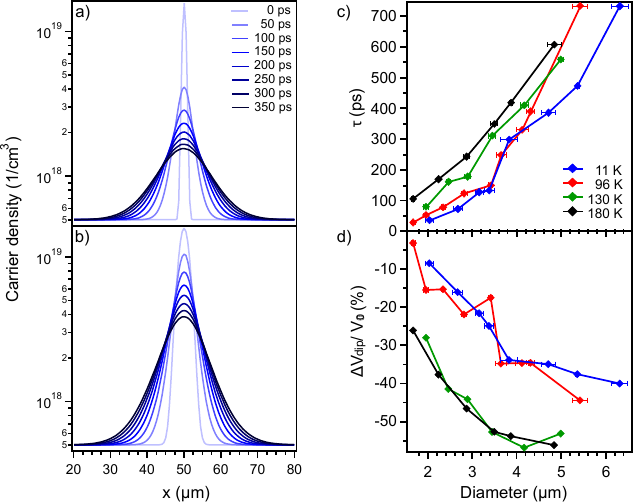}
    \caption{Laser spot diameter dependence of TPV. Simulated evolution of carrier density under a  2 $\mathrm{\mu m}$ (a) and 6 $\mathrm{\mu m}$ (b) wide Gaussian laser profile over 350 ps, respectively. Experimentally determined recovery time (c) and modulation of the dip amplitude (d) as a function of the laser spot diameter at various temperatures. Pump fluence was kept at 15 $\mu$J cm$^{-2}$, and probe at 4 $\mu$J cm$^{-2}$ for all measurements.}
\label{fig:spotsize}
\end{figure}

To further discern the TPV recovery mechanism, we varied the laser spot size by moving the device slightly off the focal plane. Since the recovery time depends on fluence, we kept the fluence constant across all spot size-dependent measurements. While the recombination time remains independent of spot size, the carrier escape time increases with larger spot sizes due to longer escape paths. More quantitatively, the simulated carrier density after a laser pulse indeed decays much slower for the larger laser spot [Figure \ref{fig:spotsize}(a,b)] (simulation details presented later). For the range of temperatures surveyed (11 - 180 K), $\tau$ increases significantly with the spot diameter [Figure \ref{fig:spotsize}(c)], supporting that the recovery is dominated by carrier escape. The magnitude of the photovoltage reduction also increases notably with spot diameter [Figure \ref{fig:spotsize}(d)], in line with the greater recombination loss as carriers escape more slowly from a larger laser spot. 

The slower recovery at higher fluence then indicates carrier diffusivity is greatly reduced at high carrier density. On the probe first side, $\tau$ is substantially shorter than that on the pump first side [Figure \ref{fig:intro}(e)]. This is reasonable as the carriers injected by the probe at low fluence diffuse faster, resulting in a faster recovery. Quantitatively, we model the photovoltage recovery process based on carrier recombination and escape. The electron concentration at the photoexcitation position follows the continuity equation \cite{vogt2020ultrafast},

\begin{equation}
    \centering
    \frac{\partial n}{\partial t} = G(t) - k_1 (n-n_0) - k_2 n (n-n_0) - \frac{n-n_0}{\tau_e(n)} 
    \label{eq:rate_sim}
\end{equation}

\noindent where $G(t)$ is the pulsed photogeneration, $n_0$ is the dark electron density, and $\tau_e(n)$ is the carrier density dependent escape time. For simplicity, we ignored the Auger term. The bimolecular recombination term is simplified from $k_2(np-n_i^2)$, with the assumptions that $n_0>>p_0$ in the $n$-type device and the photoexcitation generates electrons and holes equally. Here, $n_i$ is the intrinsic electron density and $p_0$ is the dark hole density. The photovoltage at any time is then calculated from the escape rate $(n-n_0)/\tau_e$. The photovoltage measured by the lock-in amplifier is given by the difference in time-averaged photovoltages with and without the pump, i.e., $V=\langle V(t)\rangle_{pump+probe}-\langle V(t)\rangle_{pump}$. 

\begin{figure}
    \centering
    \includegraphics[width=0.65\linewidth]{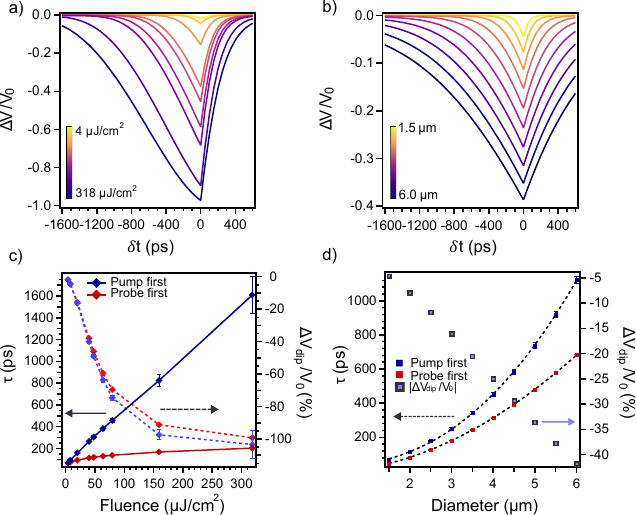}
    \caption{Modeling results of TPV. (a-b)  Simulated TPV as a function of $\delta$t at fluences and laser spot diameters similar to the experimental conditions. (c-d)  Pump fluence and spot diameter dependent $\tau$ and $\Delta V_{dip} / V_0$, extracted from exponential fittings to (a) and (b) respectively. (a) and (c) are simulated with a fixed laser radius of 1 $\mu$m; (b) and (d) with a fixed pump fluence of 5 $\mu$J cm$^{-2}$ and a probe fluence of 1 $\mu$J cm$^{-2}$. }
\label{fig:model}
\end{figure}

The simulation agrees with the experimental results well [compare Figures \ref{fig:model}(a,c) and \ref{fig:intro}(e,f)]. It produces all the main experimental features, including the larger photovoltage suppression and longer recovery time at higher fluence, as well as the faster recovery on the probe first side. To account for the strong fluence dependence, we used an escape time that scales with carrier density following a power law [$\tau_e(n) = \tau_{e0}(n/n_0)^a (n_{pump}/n_0)^b$], where $\tau_{e0}$ = 50 ps, $n_0$ = 10$^{19}$ cm$^{-3}$, $k_2$ = 10$^{-11}$ cm$^{3}$/s, $n_{pump}$ is the pump injected carrier density, $a$ = 0.85, and $b$ = 0.3. These parameters were chosen to best match our experimental results and are consistent with field effect characteristics and previously measured recombination rates \cite{gross2021nanosecond}. Other simulation parameters can be found in Table S2. The last term $(n_{pump}/n_0)^b$ is to account for the pump fluence dependence of $\tau$ on the probe first side. It might initially seem counterintuitive that the recovery time depends on the pump fluence, even when the probe arrives before the pump. We attribute this to light-induced carrier trapping \cite{hou2020nonlocal}. The pump-induced charge trapping can persist over 12.5 ns, and slow the escaping of the carriers excited by the probe pulse in the next cycle. This long-term effect is evidenced by a shift of photovoltage baseline when blocking the pump pulse (Figure S6). 

We also simulated the photovoltage suppression at various laser spot radii by setting $\tau_e (r) = \tau_{e0} (r/r_0)^2$, where the quadratic dependence is expected for diffusive transport. The simulation results are also in good agreement with the experiment [compare Figures \ref{fig:model}(b, d) and \ref{fig:spotsize}(c,d)].

\begin{figure}
    \centering
    \includegraphics[width=0.65\linewidth]{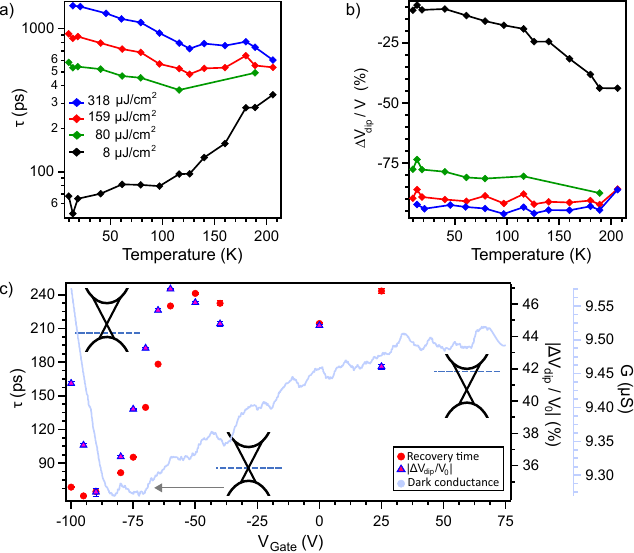}
    \caption{Temperature and gate dependence of recovery dynamics at various pump fluences. (a) Recovery time of device D1 as a function of temperature (13 K - 220 K) at various pump fluences. (b) Temperature dependence of the photovoltage suppression percentage at zero delay. (c) Photovoltage recovery time and zero-delay suppression percentage as a function of gate voltage for device D2 at 28 K (Pump: 10.7 $\mu$J cm$^{-2}$ , probe: 3.3 $\mu$J cm$^{-2}$) . Band diagrams illustrate the gate tuning of the chemical potential across the Dirac point. The light gray curve represents the dark conductance, with values shown on the second right y-axis. All $\tau$ and $|\Delta V/V_0|$ values are extracted from the pump first side.}
    \label{fig:Temp&gate}
\end{figure}

Next, we examine the temperature and gate effects on the photovoltage recovery. At low pump fluence (8.0 $\mu$J cm$^{-2}$), an exponential function provided a good fit to the photovoltage suppression curves across all tested temperatures (Figure S7). The recovery time increases by about 5 times from 13 to 200 K [Figure \ref{fig:Temp&gate}(a)], indicating slower carrier diffusion at higher temperatures. At fluence above 80 $\mu$J cm$^{-2}$, $\tau$ becomes less sensitive to temperature. $\tau$  slightly decreases at higher temperatures, opposite to the trend at low fluence. The high fluence behavior can be understood by considering the laser heating and hot carriers. First, the average sample temperature is estimated to increase by about 18 K at 80 $\mu$J cm$^{-2}$ (72 K at 320 $\mu$J cm$^{-2}$), assuming the main heat dissipation is through the SiO$_2$ layer. Second, in ultrafast measurements, the electron temperature can be significantly higher than the lattice temperature. The elevated electron temperature may make the diffusivity insensitive to the lattice temperature. The slight reduction in $\tau$ at higher temperatures may be due to carrier recombination occurring faster than escape, which contributes to the photovoltage recovery. 

Both recovery time and photovoltage suppression percentage reach their minima, when the gate voltage adjusts the conductance to its lowest value [Figure \ref{fig:Temp&gate}(c)]. This indicates that carriers diffuse faster when the chemical potential is tuned close to the Dirac point. More exactly, the minimum $\tau$ and $|\Delta V/V|$ occur at $V_g \approx$ -90 V, while the minimum conductance is at $V_g \approx$ -80 V. The slight offset may be caused by the inhomogeneity of gate-induced carrier density across the thickness of the nanoribbon. The photons are absorbed close to the top surface, where the gate-induced chemical potential shift is less effective. At $V_g >$ -60 V, $\tau$ plateaus, likely because the screening of the gate field by the remaining bulk carriers prohibits chemical potential tuning at the top surface.

\begin{figure}
    \centering
    \includegraphics[width=1\linewidth]{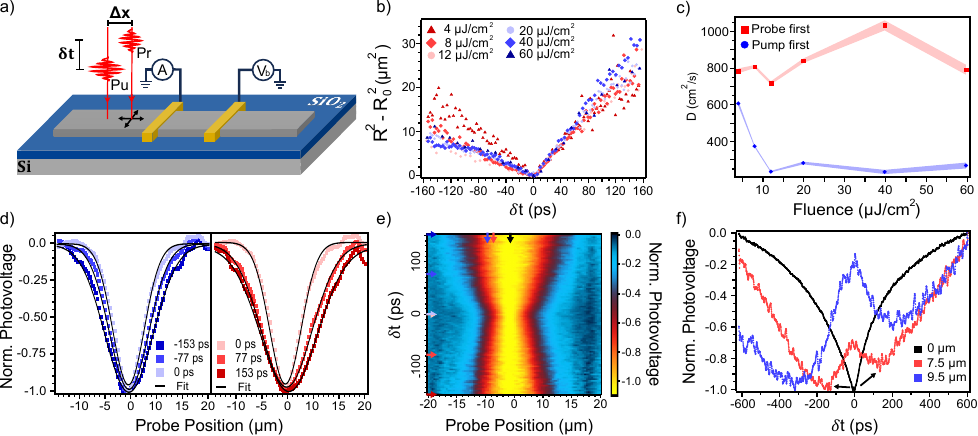}
    \caption{TPVM measurements at 21 K on device D2 incorporating a 2.7-$\mu$m-wide nanoribbon. (a) Schematic representation of the TPVM setup with the pump fixed and the probe raster scanning the sample. (b) Time evolution of R$^2$-R$_0^2$ at different pump fluences. (c) Pump fluence-dependent diffusivity extracted from the linear fitting of (b). The blue (red) data points are for pump (probe) first respectively. The shaded areas represent the uncertainty obtained from curve fitting. (e) TPVM map was obtained with a pump fluence of 8.0 $\mu$J cm$^{-2}$ and a probe fluence of 1.6 $\mu$J cm$^{-2}$. The pump position is fixed at 0 $\mu$m while the probe position varies.  (d) and (f) are horizontal and vertical cuts of the map respectively. The positions of the cuts are indicated by the arrows in (e). To obtain a better temporal resolution, (f) was obtained by scanning delay time while fixing the pump and probe positions.}
    \label{fig:spatial}
\end{figure}

Finally, we present the ultrafast transient photovoltage microscopy (TPVM) results. TPVM is based on scanning photocurrent microscopy (SPCM), which provides rich information on the local electric field distribution \cite{graham2013scanning}, carrier diffusion \cite{fu2011electrothermal,xiao2018use,mcclintock2020temperature, mcclintock2022highly}, and spatially resolved helicity-dependent photocurrent \cite{wang2023spatially,wang2024temperature}. By adding time resolution, this method is analogous to the transient absorption microscopy (TAM) \cite{guo2017long,zhou2022transient}, but here we measure the photovoltage instead of optical absorption. 

We choose to focus on a device composed of a narrower nanoribbon (device D2) showing larger photovoltage suppression in TPVM, presumably caused by the confinement of charge transport along the length. We fix the pump position and raster scan the probe beam [Figure \ref{fig:spatial}(a)]. The photovoltage is measured as a function of the probe position at each delay time. Figure \ref{fig:spatial}(e) maps the normalized photovoltage against the delay time and probe position along the nanoribbon length. The photovoltage dip exhibits the narrowest spatial waist at zero delay time. The dip is broadened when the delay time increases [Figure \ref{fig:spatial}(d)], since the carriers excited by the first pulse diffuse into a larger area. The probe first side with positive delay time shows a faster broadening, also attributed to faster diffusion of probe-induced carriers at lower density. The vertical cuts of the TPVM map yield TPV traces measured at a fixed spatial separation of pump and probe pulses [Figure \ref{fig:spatial}(f)]. When the pump and probe spatially overlap ($\Delta x$ = 0), a single dip is observed consistent with Figure \ref{fig:intro}(e). But when the pump and probe are spatially separated, two dips are clearly seen. This can be understood because the carriers injected by the first pulse must diffuse to the position of the second pulse to create nonlinear recombination. A delay time is needed to allow for carrier diffusion between the spatially separated pulses. As the spatial separation increases, the dips shift to longer delay time [Figure \ref{fig:spatial}(f)], as it takes longer for carriers to diffuse.  

To quantitatively extract the diffusivity, we fit the horizontal cross section in Figure \ref{fig:spatial}(e) at each delay time by a Gaussian function, $\Delta V = \Delta V_{dip} \exp[-x^2/(2R^2)]$, where $R_0$ is the width at zero delay. $R^2-R_0^2$ is linear with $\delta t$ [Figure \ref{fig:spatial}(b)], indicating carriers transport diffusively. As the nanoribbon being studied here has a much smaller width and thickness compared to its length, the diffusion is mainly confined to one dimension. We can then find the diffusivity by $D = (R^2-R_0^2)/2 \delta t$. On the pump first side, $D$ decreases from 600 to 250 cm$^2$/s as the pump fluence increases [Figure \ref{fig:spatial}(c)]. On the probe first side, $D$ reaches about 800 cm$^2$/s and is insensitive to pump fluence. The higher $D$ value is again attributed to the lower carrier density induced by the probe. We performed the TPVM measurements at various temperatures and the super-diffusion behavior was maintained up to 180 K (Figure S8). The TPVM results can also be simulated using the previous model based on the continuity equation, shown in Figure S9.
 
The highest reported electron mobility in bulk crystals of $\mathrm{Bi_2Se_3}$ at low temperatures is about $10^4$ cm$^2$/Vs \cite{butch2010strong}, corresponding to a diffusivity of 10 cm$^2$/s at 10 K. The CVD-grown nanoribbons typically exhibit lower electron mobility compared to bulk crystals \cite{kunakova2021high}. Hence the diffusivity values up to 800 cm$^2$/s as measured by TPVM are two or three orders of magnitude higher than that of band-edge carriers. The most obvious explanation for the high diffusivity is via hot carriers. Hot carrier diffusivity can be orders of magnitude higher than band-edge carrier diffusivity since hot carriers experience much higher temperatures than the surrounding lattice. However, our results suggest hot carriers maintain high diffusivity for hundreds of ps and travel across up to 10 $\mu$m before thermalization. Though long-lived hot carriers have been reported in Si \cite{najafi2017super} and halide perovskites \cite{yang2016observation, guo2017long}, the thermalization in TIs has been reported to be rather fast, within a few ps \cite{sobota2012ultrafast,mori2023spin}. Furthermore, hot carrier diffusivity should increase with higher laser fluence \cite{najafi2017super}, opposite to our observation. 

An alternative mechanism that accounts for the observed super-diffusion is via exciton condensation. At low temperatures, a large fraction of excitons may condense into the lowest quantum state in analogy to helium superfluids, resulting in much more efficient charge transport. Hence the high diffusivity is then linked to the exciton formation. The TPV results, including fluence, temperature, and gate dependence, are all consistent with this mechanism. The reduced diffusivity at higher fluence results from enhanced electric screening, which weakens exciton formation at high carrier densities. Higher temperatures inhibit exciton formation, which slows both carrier escape and recovery processes. Diffusion becomes faster when the device is tuned intrinsic by gate, enabling efficient exciton formation. The TPVM results indicate diffusive transport, while the exciton condensate is expected to travel ballistically. This dependency can be understood by considering that the laser fluence used in the experiment is too high for complete exciton formation. In this case, the photogenerated carriers dynamically convert between free carriers and excitons \cite{tang2021transport}. The free carriers suffer from scattering, leading to apparent diffusive transport. 

In summary, we have observed super-diffusion of photoexcited charge carriers in TI nanoribbons. The transient photovoltage is substantially suppressed by a pump pulse caused by the nonlinear recombination rate. Laser spot size dependent measurements demonstrate that the photovoltage recovery is dominated by the carrier diffusion out of the laser spot. A simple diffusion-recombination model with a carrier density dependent diffusivity agrees well with the experimental results. The carrier diffusivity is two or three orders of magnitude higher than that expected from band-edge free carriers. The diffusivity is highest in intrinsic devices and is reduced at high carrier density and high temperatures. These results do not fully agree with the behavior expected from the hot carriers but are consistent with the formation of exciton condensates in TIs. The developed TPV and TPVM methods can be widely used in a variety of materials to study their ultrafast carrier relaxation and transport properties.

\section{Methods}

Single-crystalline Sb-doped Bi$_2$Se$_3$ nanoribbons were synthesized following a previous report \cite{hou2019millimetre}. 125 mg of Bi$_2$Se$_3$ granules (99.999\% Thermo Scientific) mixed with 40 mg of Sb powder (99.999\% Alfa Aesar) were placed in the center of a quartz tube furnace. 250 mg of Se pellets (99.999\% Johnson Matthey Inc.) were placed 36 cm upstream from the center and a 10 nm gold-coated Si substrate was placed 8 cm downstream from the center. Next, the quartz tube was evacuated until a base pressure of around 50 mTorr was reached and then brought back to atmosphere with Ar gas. The center of the furnace was brought up to 660 $^{\circ}$C with a 150 sccm Ar flow. The growth lasted 3.5 hrs and was cooled naturally.  After the growth, the nanostructures were mechanically transferred on top of degenerately $n$-doped silicon substrates coated by 300 nm silicon dioxide. Top contacts were patterned with electron beam lithography (EBL). A 5 nm chromium adhesion layer followed by a 295 nm gold layer were e-beam evaporated. An optical image of a final device is shown in Figure \ref{fig:intro}(c). 

To perform TPV, a Ti:Sapphire laser (Mai Tai) with 200 fs pulse width and 80 MHz repetition rate was separated at the beam splitter (BS) and the probe beam was modulated by a photoelastic modulator (PEM)-based chopper  [Figure~ \ref{fig:intro}(b)]. The probe can be scanned on the device plane by scan mirrors to obtain spatial photovoltage maps. The pump line was passed through a delay stage (Newport) to control the delay time. The pump and probe beams were cross-polarized to avoid any interference. The beams were then reunited with another BS and focused through the objective lens (40$\times$ N.A.=0.6) onto the sample placed in a cryostat (Janis ST-500). A lock-in amplifier was used to measure probe-induced photovoltage as a function of delay time. For spot size dependence, the spot diameter was measured using a knife edge technique from the reflectance change as the laser spot scans across a sharp metal edge (Figure S10).

\begin{acknowledgement}

This work was supported by the U.S. National Science Foundation Grants No. DMR-2404957 and No. DMR-2209884. Part of this study was performed at the UC Davis Center for Nano and Micro Manufacturing (CNM2). We acknowledge P. Klavins for his assistance in maintaining the helium recovery and liquefaction system, and T. Valenti, A. Badrinath for assisting with the experiments and simulation.

\end{acknowledgement}

\begin{suppinfo}

Electronic characterization of TI nanoribbon devices, additional photocurrent images, TPV results, modeling, and experimental details. 

\end{suppinfo}


\providecommand{\latin}[1]{#1}
\makeatletter
\providecommand{\doi}
  {\begingroup\let\do\@makeother\dospecials
  \catcode`\{=1 \catcode`\}=2 \doi@aux}
\providecommand{\doi@aux}[1]{\endgroup\texttt{#1}}
\makeatother
\providecommand*\mcitethebibliography{\thebibliography}
\csname @ifundefined\endcsname{endmcitethebibliography}  {\let\endmcitethebibliography\endthebibliography}{}

\end{document}